\documentclass[sigconf, anonymous=False]{acmart}
\usepackage{tabularx, booktabs}
\usepackage{graphicx}
\usepackage{multirow}
\usepackage{bm}
\usepackage{balance}
\usepackage{enumitem}
\usepackage{caption}
\usepackage{diagbox}
\usepackage{booktabs}
\usepackage{subfig}
\usepackage{siunitx}
\usepackage{footmisc}
\usepackage{footnote}
\usepackage[flushleft]{threeparttable}
\usepackage{etoolbox}
\usepackage{dsfont}
\appto\TPTnoteSettings{\footnotesize}

\usepackage{tablefootnote}
\makesavenoteenv{table}
\makesavenoteenv{tabular}

\newcolumntype{C}[1]{>{\centering\arraybackslash}p{#1}}
\newcolumntype{R}[1]{>{\raggedleft\arraybackslash}p{#1}}
\newcolumntype{L}[1]{>{\raggedright\arraybackslash}p{#1}}

\newcommand\mc[1]{\multicolumn{1}{c}{#1}}

\newcommand\blfootnote[1]{%
\begingroup 
\renewcommand\thefootnote{}\footnote{#1}%
\addtocounter{footnote}{-1}%
\endgroup 
}

\makeatletter
\newcommand{\ssymbol}[1]{^{\@fnsymbol{#1}}}
\makeatother

\makeatletter
\def\hlinew#1{%
  \noalign{\ifnum0=`}\fi\hrule \@height #1 \futurelet
   \reserved@a\@xhline}
\makeatother

\AtBeginDocument{%
  \providecommand\BibTeX{{%
    \normalfont B\kern-0.5em{\scshape i\kern-0.25em b}\kern-0.8em\TeX}}}

\acmConference[WSDM'22]{The 15th ACM International Conference on Web Search and Data Mining}{February 21-25, 2022}{Phoenix, Arizona}

\begin{document}

\title{Learning Discrete Representations via Constrained Clustering for Effective and Efficient Dense Retrieval}


\author{Jingtao Zhan$^{1}$, Jiaxin Mao$^{2}$, Yiqun Liu$^{1\star}$, Jiafeng Guo$^{3}$, Min Zhang$^{1}$, Shaoping Ma$^{1}$}
\affiliation{%
  \institution{${1}$ Department of Computer Science and Technology, Institute for Artificial Intelligence, \\
  Beijing National Research Center for Information Science and Technology, Tsinghua University, Beijing 100084, China
  }
  }
  
\affiliation{%
  \institution{${2}$ Beijing Key Laboratory of Big Data Management and Analysis Methods, Gaoling School of Artificial Intelligence, \\ Renmin University of China, Beijing 100872, China
}
  }

\affiliation{%
  \institution{${3}$ CAS Key Lab of Network Data Science and Technology, Institute of Computing Technology, \\
  		Chinese Academy of Sciences, Beijing, China
  }
  }

\email{jingtaozhan@gmail.com, maojiaxin@gmail.com, yiqunliu@tsinghua.edu.cn, guojiafeng@ict.ac.cn}

\renewcommand{\shortauthors}{Zhan, et al.}

\begin{abstract}
Dense Retrieval~(DR) has achieved state-of-the-art first-stage ranking effectiveness. However, the efficiency of most existing DR models is limited by the large memory cost of storing dense vectors and the time-consuming nearest neighbor search~(NNS) in vector space. Therefore, we present RepCONC, a novel retrieval model that learns discrete Representations via CONstrained Clustering. RepCONC jointly trains dual-encoders and the Product Quantization~(PQ) method to learn discrete document representations and enables fast approximate NNS with compact indexes. It models quantization as a constrained clustering process, which requires the document embeddings to be uniformly clustered around the quantization centroids and supports end-to-end optimization of the quantization method and dual-encoders. We theoretically demonstrate the importance of the uniform clustering constraint in RepCONC and derive an efficient approximate solution for constrained clustering by reducing it to an instance of the optimal transport problem. Besides constrained clustering, RepCONC further adopts a vector-based inverted file system~(IVF) to support highly efficient vector search on CPUs. Extensive experiments on two popular ad-hoc retrieval benchmarks show that RepCONC achieves better ranking effectiveness than competitive vector quantization baselines under different compression ratio settings. It also substantially outperforms a wide range of existing retrieval models in terms of retrieval effectiveness, memory efficiency, and time efficiency.
\end{abstract}

\ccsdesc[500]{Information systems~Search index compression}
\ccsdesc[500]{Information systems~Retrieval models and ranking}
\ccsdesc[300]{Information systems~Information retrieval}

\keywords{index compression, dense retrieval, neural ranking}

\maketitle

\blfootnote{$^\star$Corresponding author}

\section{Introduction}

Dense Retrieval (DR) has become a popular paradigm for first-stage retrieval in ad-hoc retrieval tasks. Through embedding queries and documents in a latent vector space with dual-encoders and using nearest neighbor search to retrieve relevant documents, the DR paradigm avoids the vocabulary mismatch problem, which has been a great challenge for traditional bag-of-words~(BoW) models~\cite{robertson1994some}. With end-to-end supervised training, recent works have achieved state-of-the-art ranking performance and significantly outperforms BoW models~\cite{lin2020distilling, qu2021rocketqa, xiong2021approximate, zhan2020repbert}.

Despite the success in improving ranking performance, most existing DR models~\cite{zhan2021optimizing, xiong2021approximate, qu2021rocketqa, karpukhin2020dense} are inefficient in memory usage and computational time. 
For memory inefficiency, the size of the embedding index is usually an order of magnitude larger than that of BoW index~\cite{zhan2021jointly}. At runtime, the vectors must be loaded to system memory or even GPU memory, which is both costly and highly limited in size. 
As for time inefficiency, many existing DR models~\cite{zhan2021optimizing, xiong2021approximate, qu2021rocketqa, karpukhin2020dense} do not use approximate vector search~\cite{jegou2010product, malkov2018efficient}. They have to conduct exhaustive search, i.e., computing relevance scores between the submitted query and all documents, which is less time-efficient than BoW models with inverted indexes. As a result, these DR models cannot use CPUs for retrieval due to high latency and have to use much more expensive GPUs to accelerate the search~\cite{zhan2021jointly, zhan2021optimizing, xiong2021approximate}. 

A key solution for the efficiency issue of DR models is to learn discrete representations for document embeddings, which can be encoded into compact indexes and enable efficient vector search. Popular methods for learning this kind of discrete representation include Product Quantization~(PQ)~\cite{jegou2010product, ge2013optimized} and Locality Sensitive Hashing~(LSH)~\cite{indyk1998approximate}. However, these methods usually learn discrete representations in an unsupervised way and cannot benefit from supervised signals. Directly adopting these techniques usually hurts ranking effectiveness~\cite{zhan2021jointly, zhang2021joint}. 

Therefore, jointly optimizing dual-encoders and the quantization methods with supervised labels is regarded as a promising direction in improving retrieval effectiveness. 
However, it is inherently challenging because the quantization operation is non-differentiable and the model cannot be trained in an end-to-end fashion. There exist a number of recent works (e.g., JPQ~\cite{zhan2021jointly}) trying to solve this problem, but they usually suffer from significant performance loss while improving efficiency. Therefore, we believe it is still an unsolved but essential problem. 

To tackle this problem, we present RepCONC, which stands for learning discrete \textbf{Rep}resentations via \textbf{CON}strained \textbf{C}lustering\footnote{Code and models are available at \url{https://github.com/jingtaozhan/RepCONC}.}. It jointly trains the dual-encoders and PQ by modeling quantization as a \emph{constrained clustering} process. Specifically, \emph{constrained clustering} involves a clustering loss and a uniform clustering constraint. 
The clustering loss is introduced to train the discrete codes with the requirement that document embeddings are clustered around the quantization centroids. 
We also employ a uniform clustering constraint, which requires the vectors to be equally assigned to all quantization centroids. 
We add the constraint because we find that unconstrained clustering tends to assign vectors to a few major clusters and makes the quantized vectors indistinguishable with each other.  
Since this constraint leads to a difficult combinatorial optimization problem, we derive an approximate solution by relaxing it to an instance of the optimal transport problem. 
Besides the two components of \emph{constrained clustering}, RepCONC further employs vector-based inverted file system (IVF)~\cite{jegou2010product}, which enables efficient non-exhaustive vector search. 
With these designs, RepCONC can run on either GPU or CPU~\footnote{Except that the user queries are still encoded on GPU.} and perform vector search in an efficient way.  

We conduct experiments on two widely-adopted ad-hoc retrieval benchmarks~\cite{bajaj2016ms, craswell2020overview} and compare RepCONC with a wide range of baselines, including both vector compression methods and retrieval models. Experimental results show that: 1) RepCONC significantly outperforms competitive vector compression baselines with different compression ratio settings from tens of times to hundreds of times. 2) RepCONC substantially outperforms various retrieval baselines in terms of retrieval effectiveness, memory efficiency, and time efficiency. 3) The ablation study demonstrates that \emph{constrained clustering} is the key to the effectiveness of RepCONC.

\section{Related Works}

DR represents queries and documents with embeddings and utilizes vector search to retrieve relevant documents. Most existing DR models~\cite{karpukhin2020dense, zhan2020repbert, xiong2021approximate, zhan2021optimizing, qu2021rocketqa} share the same BERT-base~\cite{devlin2019bert, liu2019roberta} architecture and utilize brute-force vector search. They differ in training methods, which can be classified into two categories. One line of research is negative sampling~\cite{huang2020embedding, karpukhin2020dense, zhan2020repbert, xiong2021approximate, zhan2021optimizing}. According to \citet{zhan2021optimizing}, utilizing hard negatives helps improve top ranking performance. The other line is knowledge distillation~\cite{qu2021rocketqa, lin2020distilling, hofstatter2021efficiently}, which adopts a cross-encoder to generate pseudo labels. This paper uses negative sampling to train RepCONC and leaves training RepCONC with knowledge distillation to future work. 

Since the models mentioned above utilize brute-force vector search, they incur very large embedding indexes and have to use costly GPUs to accelerate the search. How to address the efficiency issue has recently attracted researchers' attention. Several studies propose some workarounds~\cite{yamada2021bpr, zhang2021joint, zhan2021jointly}. 
BPR~\cite{yamada2021bpr} binarizes dense vectors and is conducted on the OpenQA task. An obvious limitation is that the compression ratio is fixed to 32x. DPQ~\cite{zhang2021joint, chen2020differentiable} utilizes PQ~\cite{jegou2010product} for compression and is designed for word embedding compression and recommendation systems. Most recently, \citet{zhan2021jointly} propose JPQ for document ranking and achieve state-of-the-art results. JPQ utilizes fixed discrete codes~(Index Assignments) generated by K-Means and only trains the query encoder and PQ Centroid Embeddings. 
RepCONC is different from JPQ in both joint learning framework and efficiency design. Firstly, with the help of constrained clustering, RepCONC is able to optimize discrete codes~(Index Assignments) while JPQ cannot. Secondly, RepCONC additionally employs the inverted file system~(IVF)~\cite{jegou2010product} to accelerate search and thus, can efficiently retrieve documents on CPUs while JPQ has to rely on GPUs.

\section{Constrained Clustering Model}

In this section, we propose RepCONC, which stands for learning discrete \textbf{Rep}resentations via \textbf{CON}strained \textbf{C}lustering. We firstly introduce the preliminary of Production Quantization~\cite{jegou2010product}, a widely-used vector compression method for approximate nearest neighbor search~(ANNS). Then we elaborate our model.

\subsection{Revisiting Product Quantization}
\label{sec:revisit_pq}
RepCONC is based on Product Quantization~(PQ)~\cite{jegou2010product}. For vectors of dimension $D$, PQ defines $M$ sets of embeddings, each of which includes $K$ embeddings of dimension $D/M$. They are called PQ Centroid Embeddings. 
Formally, let $\bm{c}_{i,j}$ be the $j_{th}$ centroid embedding from the $i_{th}$ set:
\begin{equation}
	\bm{c}_{i,j} \in \mathbb{R}^\frac{D}{M} \quad (1 \leq i \leq M, 1 \leq j \leq K) 
\end{equation}
Given a document embedding $\bm{d} \in \mathbb{R}^{D}$, PQ firstly splits it into $M$ sub-vectors.
\begin{equation}
	 \bm{d} = \bm{d}_1,\bm{d}_2,...,\bm{d}_M
\end{equation}
Then PQ independently quantizes each sub-vector to the nearest PQ Centroid Embedding. Formally, to quantize a sub-vector $\bm{d}_i$, PQ selects $\bm{c}_{i, \varphi_i(d)}$ which achieves the minimum quantization error:
\begin{equation}
\label{eq:select_index_with_argmin}
	\varphi_i(d) = \arg \min_{j} \lVert \bm{c}_{i, j} - \bm{d}_i \rVert ^2
\end{equation}
Let $\bm{\varphi}(d)$ be the concatenation of $\varphi_i(d)$:
\begin{equation}
	\bm{\varphi}(d) = \varphi_1(d),\varphi_2(d), ..., \varphi_i(M) \in \{1,2,..., K\}^M
\end{equation}
where comma denotes vector concatenation. $\bm{\varphi}(d)$ is called the Index Assignment of $d$. Along with the PQ Centroid Embeddings, $\bm{\varphi}(d)$ can reconstruct the quantized document embedding $\bm{\hat{d}}$ as follows:
\begin{equation}
	\bm{\hat{d}} = \bm{c}_{1,\varphi_1(d)},\bm{c}_{2,\varphi_2(d)},...,\bm{c}_{M,\varphi_M(d)} \in \mathbb{R}^D
\end{equation}

PQ improves both memory efficiency and time efficiency.
For memory efficiency, PQ does not explicitly store $\bm{d}$ or $\bm{\hat{d}}$. Instead, it only stores the PQ Centroid Embeddings $\{\bm{c}_{i,j}\}$ and Index Assignments $\bm{\varphi}(d)$. 
Since $K$ is usually less than or equal to 256, $\bm{\varphi}(d)$ can be encoded with $M$ bytes. Therefore, the compression ratio is about $4D/M$.
As for time efficiency, PQ enables efficient vector search. Given a query embedding, PQ splits it equally to $M$ sub-vectors and pre-computes the similarities between the sub-vectors and PQ Centroid Embeddings. Then, PQ efficiently computes the similarities between the query embedding and each document embedding by aggregating the corresponding pre-computed similarities. Compared with directly computing vector similarity, the speedup ratio is about $D/M$. 

\subsection{Clustering and Representation Learning}

\begin{figure}
    \includegraphics[width=0.95\linewidth, keepaspectratio=True]{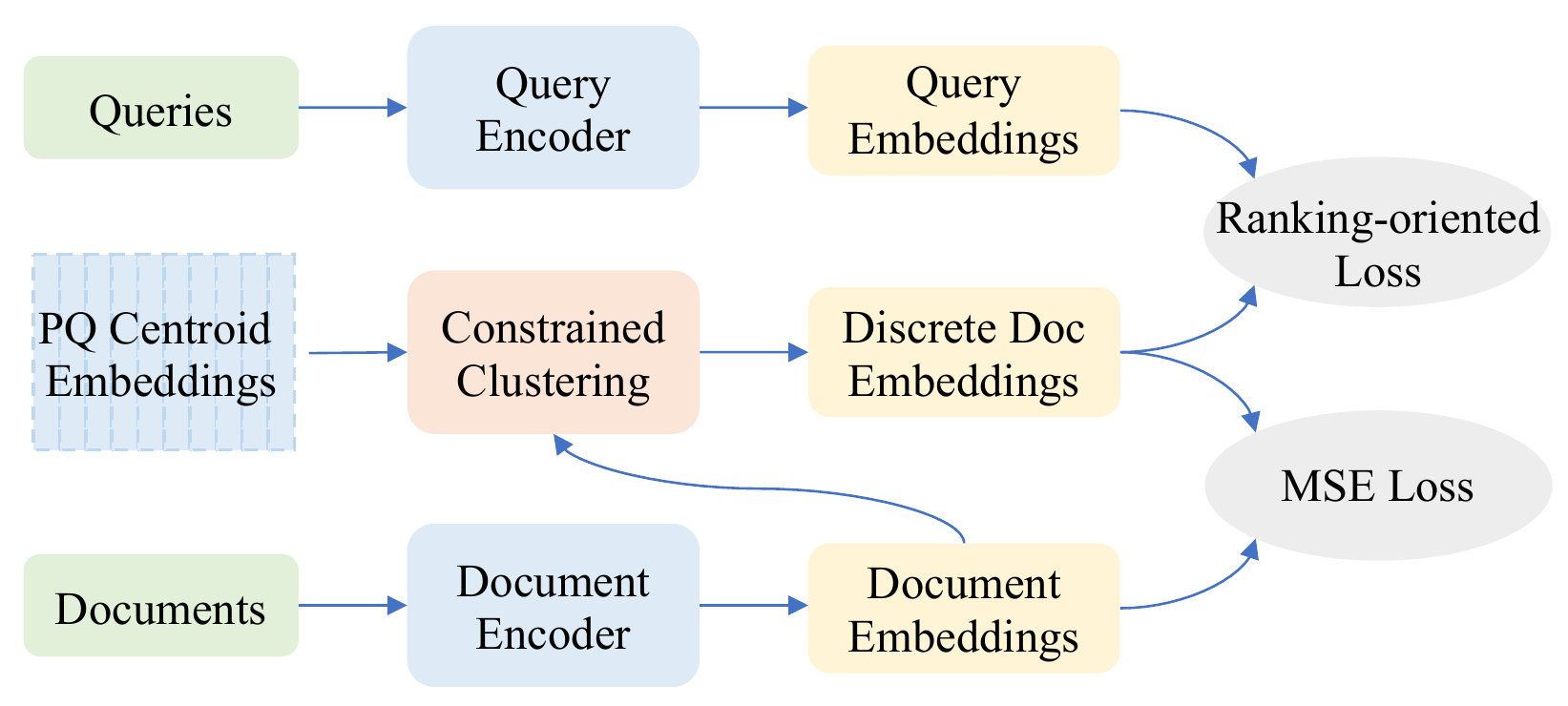}
    \caption{Training process of RepCONC.} 
    \label{fig:workflow_repconc}
\end{figure}

Jointly optimizing dual-encoders and PQ parameters is challenging. RepCONC views it as a simultaneous clustering and representation learning problem. To solve it, RepCONC utilizes both the ranking-oriented loss~\cite{zhan2021jointly} and a clustering loss for training. The ranking-oriented loss helps learn representations for ranking. The clustering loss, i.e., the mean square error (MSE) between the document embeddings and the quantization centroids, helps cluster document embeddings to centroid embeddings. We illustrate the training workflow in Figure~\ref{fig:workflow_repconc}.

The ranking-oriented loss~\cite{zhan2021jointly} replaces the uncompressed document embeddings in the common DR ranking loss functions~\cite{qu2021rocketqa, xiong2021approximate, zhan2021optimizing} with the quantized document embeddings. Therefore, it better evaluates the ranking performance with respect to the current compression parameters. Formally, ranking-oriented loss $L_{r}$ is formulated as:
\begin{equation}
\label{eq:ranking_oriented_loss}
\begin{aligned}
L_{r} &= - \log \frac{\mathrm{e}^{\langle \bm{q}, \bm{\hat{d}}^+ \rangle}}{\mathrm{e}^{\langle \bm{q}, \bm{\hat{d}}^+ \rangle} + \sum_{d^-} \mathrm{e}^{\langle \bm{q}, \bm{\hat{d}}^- \rangle}} \\
\end{aligned}
\end{equation}
where $d^+$ and $d^-$ are relevant and irrelevant documents, respectively. $L_{r}$ facilitates effective representation learning by encouraging the relevant pairs to be scored higher than irrelevant pairs. 

Although incorporating the ranking-oriented loss helps a recent joint learning work achieve state-of-the-art compression results~\cite{zhan2021jointly}, we argue that simply relying on this loss is problematic. Since the quantization error in Eq.~(\ref{eq:select_index_with_argmin}) is not included in $L_{r}$, it may change arbitrarily and selecting Index Assignments based on Eq.~(\ref{eq:select_index_with_argmin}) may lead to unexpected behaviors. \citet{zhan2021jointly} avoids this pitfall by fixing the Index Assignments. However, fixed Index Assignments lead to sub-optimal ranking performance because they cannot benefit from supervised signals. 

Different from \citet{zhan2021jointly}, RepCONC regards quantization as a clustering problem and introduces the MSE loss $L_{m}$:
\begin{equation}
\begin{aligned}
L_{m} &= \lVert \bm{d} - \bm{\hat{d}} \rVert ^2
\end{aligned}
\end{equation}
Minimizing $L_{m}$ requires the vectors before and after quantization to be close to each other. In this way, the document embeddings are clustered around the centroid embeddings. Combining both $L_{r}$ and $L_{m}$ helps the model to cluster document embeddings based on ranking effectiveness. It is expected to produce better clustering compared with unsupervised training. 

The final loss $L$ is a weighted sum of ranking-oriented loss $L_{r}$ and the MSE loss $L_{m}$.
\begin{equation}
\label{eq:weighted_sum_loss}
\begin{aligned}
L &= L_{r} + \lambda L_{m}
\end{aligned}
\end{equation}
$\lambda$ is a hyper-parameter. If $\lambda$ is too small, the documents are not clustered and the selected Index Assignments become arbitrary. If $\lambda$ is too big, the MSE loss dominates the training process and harms ranking effectiveness. 
In practice, we find the model is relatively sensitive to $\lambda$, but becomes more robust and effective with the help of the uniform clustering constraint introduced in the next sections.

Since quantization involves some non-differentiable operations, we explicitly design the gradient back-propagation policy.
The gradients of uncompressed document embeddings are defined as follows:
\begin{equation}
\begin{aligned}
\frac{\partial L}{\partial \bm{d}} & \coloneqq \frac{\partial L_{r}}{\partial \bm{\hat{d}}} + \lambda \frac{\partial L_{m}}{\partial \bm{d}} \\
\end{aligned}
\end{equation}
As the equation shows, we add the gradient of quantized document embeddings~(the first term). The gradients are further back-propagated to dual-encoders.
As for the PQ Centroid Embeddings, their gradients can be derived with chain rule. We formally show it as follows:
\begin{equation}
\label{eq:grad_for_pq_centroids}
\left\{
  \begin{array}{clr}
      \frac{\partial L}{\partial \bm{\hat{d}}} &= \frac{\partial L_{r}}{\partial \bm{\hat{d}}} + \lambda \frac{\partial L_{m}}{\partial \bm{\hat{d}}} \\[10pt]
       \frac{\partial L}{\partial \bm{c}_{i,j}} &= \bm{1}_{\varphi_i(d)=j} \cdot \frac{\partial L}{\partial \bm{\hat{d}}}  \\
  \end{array}
\right.
\end{equation}
In the following sections, we show how RepCONC selects Index Assignments~(instead of Eq.~(\ref{eq:select_index_with_argmin})).

\subsection{Importance of Uniform Clustering}

\begin{figure}
    \includegraphics[width=0.95\linewidth, keepaspectratio=True]{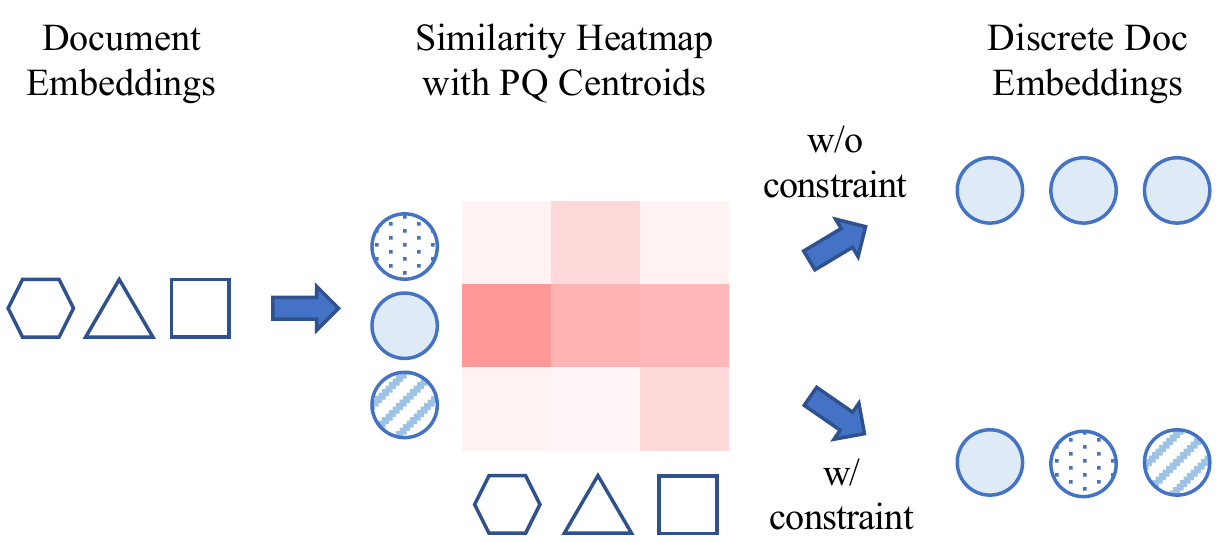}
    \caption{Illustration of Constrained Clustering. Darker colors in the heatmap indicate higher similarities~(smaller distances). With the constraint, the discrete document embeddings are more diverse. } 
    \label{fig:constrain_cluster}
\end{figure}

It is non-trivial to simultaneously conduct clustering and representation learning because the two objects are conflicting to some extent. 
Although representation learning encourages vectors to be distinguishable, clustering encourages vectors to be identical. In practice, clustering tends to map vectors to several major clusters while some clusters are rarely used or even empty. The problem worsens with Eq.~(\ref{eq:grad_for_pq_centroids}) where rarely used centroid embeddings are less likely to be updated and may end up with arbitrary values. 
The unbalanced clustering distribution affects the distinguishability of the quantized vectors and compromises ranking effectiveness.

We tackle this challenge by imposing a uniform clustering constraint. It requires the document sub-vectors to be equally assigned to all PQ Centroid Embeddings. The learning object along with the constraint is formally expressed as:
\begin{equation}
\begin{aligned}
\min L \quad \text{subject to } \forall i,j: P(\varphi_i(d)=j) = \frac{1}{K}   \\
\end{aligned}
\end{equation}
We illustrate constrained clustering in Figure~\ref{fig:constrain_cluster}. As the figure shows, the discrete document embeddings are selected by minimizing the quantization error~(maximizing the similarity) given the uniform clustering constraint. Without the constraint, the discrete document embeddings become identical.
In the following, we theoretically analyze the importance of uniform clustering.

We introduce several notations for our theoretical analysis.
We define $\mathcal{I}$ as all possible Index Assignments:
\begin{equation}
\begin{aligned}
\mathcal{I} \coloneqq \{ \bm{\varphi}(d) \} = \{1,2,..., K\}^M \\
\end{aligned}
\end{equation}
Let $\bm{I} \in \mathcal{I}$ be one Index Assignment and $I_i \in \{1,2,..., K\}$ be the $i_{th}$ value of $\bm{I}$. 

Firstly, we show that maximizing the distinguishability of vectors is equivalent to forcing the vectors to be equally quantized to all possible Index Assignments. Let $\bm{\hat{d}}^s$ and $\bm{\hat{d}}^t$ be randomly sampled from all quantized document embeddings. We assume they are independent and identically distributed~(i.i.d). The probability that they are equal satisfies:
\begin{equation}
\begin{aligned}
P(\bm{\hat{d}}^s = \bm{\hat{d}}^t) & = P(\bm{\varphi}(d^s)=\bm{\varphi}(d^t)) \\
&= \sum_{\bm{I} \in \mathcal{I}} P(\bm{\varphi}(d^s) = \bm{I}) P(\bm{\varphi}(d^t) = \bm{I}) \\
& \stackrel{\text{i.i.d}}{=} \sum_{\bm{I} \in \mathcal{I}} P^2(\bm{\varphi}(d) = \bm{I}) \\
& \geq \frac{(\sum_{\bm{I} \in \mathcal{I}} P(\bm{\varphi}(d) = \bm{I}))^2}{|\mathcal{I}|} \\
&= \frac{1}{|\mathcal{I}|}
\end{aligned}
\end{equation}
where AM–GM inequality is used. The equality is achieved if and only if 
\begin{equation}
\label{eq:equal_quant_to_index}
\begin{aligned}
\forall \bm{I}: P(\bm{\varphi}(d) = \bm{I}) &= \frac{1}{|\mathcal{I}|} = \frac{1}{K^M}
\end{aligned}
\end{equation}
That is to say, quantizing vectors equally to all possible Index Assignments helps representations to be distinguishable.

Next, we show uniformly clustering sub-vectors is the essential condition of Eq.~(\ref{eq:equal_quant_to_index}). Given Eq.~(\ref{eq:equal_quant_to_index}), the probability that a sub-vector is quantized to a centroid is a constant:
\begin{equation}
\label{eq:uniform_cluster_derive}
\begin{aligned}
\forall i,j: P(\varphi_i(d) = j) = \sum_{\bm{I} : I_i = j} P(\bm{\varphi}(d)=\bm{I}) = \frac{|\{\bm{I} : I_i = j\}|}{K^M} = \frac{1}{K} 
\end{aligned}
\end{equation}

We further show that if sub-vectors are independent, uniformly clustering sub-vectors~(Eq.~(\ref{eq:uniform_cluster_derive})) is the sufficient condition of Eq.~(\ref{eq:equal_quant_to_index}):
\begin{equation}
\begin{aligned}
P(\bm{\varphi}(d) = \bm{I}) & = \prod_{i=1}^{M} P(\varphi_i(d) = I_i) = \frac{1}{K^M}
\end{aligned}
\end{equation}
Although independence among sub-vectors may not hold for practical dual-encoders, we believe constraining quantization with Eq.~(\ref{eq:uniform_cluster_derive}) is still helpful for distinguishing quantized vectors. 

\subsection{Constrained Clustering Optimization}

This section shows how to incorporate the uniform clustering constraint during training.
In previous works related to joint learning with PQ~\cite{zhan2021jointly, zhang2021joint, chen2020differentiable}, the Index Assignments are selected based on Eq.~(\ref{eq:select_index_with_argmin}). However, it cannot be applied to RepCONC because of the uniform clustering constraint. Next, we show how RepCONC incorporates the constraint to select Index Assignments during training. 

We introduce a posterior distribution $q(j|\bm{d}_i)$, which is the probability that the sub-vector $\bm{d}_i$ is quantized to the centroid $\bm{c}_{i,j}$. The Index Assignment, $\varphi_i(d)$, is the centroid with the maximum probability:
\begin{equation}
\label{eq:select_index_with_posterior_distrib}
	\varphi_i(d) = \arg \max_{j} q(j|\bm{d}_i)
\end{equation}
For previous works~\cite{zhan2021jointly, zhang2021joint, chen2020differentiable} that use Eq.~(\ref{eq:select_index_with_argmin}), $q(j|\bm{d}_i)$ can be regarded as being computed solely based on quantization error. Here for RepCONC, we compute $q(j|\bm{d}_i)$ by minimizing the quantization error given the uniform clustering constraint:
\begin{equation}
\label{eq:ideal_posterior_distrib_comput}
\begin{aligned}
\forall i: \min _{q} \sum_{d \in \mathcal{D}} \sum_{j=1}^{K} q(j|\bm{d}_i) \lVert \bm{c}_{i, j} - \bm{d}_i \rVert ^2 \text { subject to } \\ 
\forall j,d: q(j|\bm{d}_i) \in\{0,1\}, \sum_{j=1}^K q(j|\bm{d}_i)= 1 \text {, and } \sum_{d \in \mathcal{D}} q(j|\bm{d}_i)=\frac{|\mathcal{D}|}{K}
\end{aligned}
\end{equation}
where $\mathcal{D}$ indicates the set of all documents. The first condition constrains $q(j|\bm{d}_i)$ to be binary, the second condition is a natural requirement for probability, and the third condition is exactly the uniform clustering constraint. Without the third condition, Eq.~(\ref{eq:select_index_with_posterior_distrib}) and (\ref{eq:ideal_posterior_distrib_comput}) degenerate to Eq.~(\ref{eq:select_index_with_argmin}), i.e., selecting Index Assignments with minimum quantization error.  

Solving Eq.~(\ref{eq:ideal_posterior_distrib_comput}) is particularly difficult because it is a combinatorial optimization problem with the scale of millions or even billions of documents. Therefore, we use an approximate solution by relaxing $q$ to be continuous and focusing on uniformly clustering a mini-batch of documents $\mathcal{B}$:
\begin{equation}
\label{eq:approximate_solve_posterior_distrib_comput}
\begin{aligned}
\forall i: \min _{q} \sum_{d \in \mathcal{B}} \sum_{j=1}^{K} q(j|\bm{d}_i)  \lVert \bm{c}_{i, j} - \bm{d}_i \rVert ^2 \\ \text { subject to } 
\forall d: \sum_{j=1}^K q(j|\bm{d}_i)= 1 \text{ and } \forall j: \sum_{d \in \mathcal{B}} q(j|\bm{d}_i)=\frac{|\mathcal{B}|}{K}
\end{aligned}
\end{equation}
Since $\lVert \bm{c}_{i, j} - \bm{d}_i \rVert ^2$ can be regarded as the cost of mapping $\bm{d}_i$ to $\bm{c}_{i, j}$, this is an instance of the optimal transport problem and can be solved in polynomial time by linear program. In our implementation, we use Sinkhorn-Knopp algorithm~\cite{cuturi2013sinkhorn} to efficiently solve Eq.~(\ref{eq:approximate_solve_posterior_distrib_comput}). 


\subsection{Accelerating Search with IVF}
Besides PQ, RepCONC employs the inverted file system (IVF) to accelerate vector search. After quantizing document embeddings, RepCONC uses k-means to generate $n$ clusters. Each document embedding belongs to the nearest cluster and is stored in the corresponding inverted list. Note that $n$ is much smaller than the corpus size. Given a query embedding, RepCONC selects the nearest $\tilde{n}$ clusters and only ranks the documents in them. The documents in other clusters are ignored. In this way, RepCONC approximately accelerates vector search by $n/\tilde{n}$.

Note that RepCONC does not include IVF in the joint learning framework and simply uses IVF after training. The clusters are generated in an unsupervised manner. In practice, we find this already yields satisfying results, and thus training IVF with supervised labels is not explored.

IVF only induces negligible memory overhead and does not harm memory efficiency. For example, on MS MARCO Passage Ranking dataset~\cite{bajaj2016ms} which has $8$ million passages, $n$ is set to $5,000$ and additional memory overhead is less than $3\%$. 

\subsection{Training/Inference Details} 
\subsubsection{Warmup with OPQ}
\label{sec:warmup_details}
In order to accelerate convergence, we warmup the dual-encoders and PQ Centroid Embeddings as follows. We use the open-sourced STAR~\cite{zhan2021optimizing} model to initialize dual-encoders. STAR is trained without quantization. Given the document embeddings output by STAR, we use OPQ~\cite{ge2013optimized} to warmup PQ parameters, which is a popular unsupervised PQ variant. 

\subsubsection{Two-Stage Negative Sampling}
Hard negative sampling is shown to be important for retrieval models~\cite{zhan2021optimizing, xiong2021approximate}. Following \citet{zhan2021optimizing}, we train RepCONC in two stages. In the first stage, we retrieve static hard negatives using the initialized RepCONC. In the second stage, we use dynamic hard negatives, the top irrelevant documents retrieved at each training step. To enable end-to-end retrieval during training, we fix the Index Assignments and only train the query encoder and PQ Centroid Embeddings.

\subsubsection{Efficient Encoding during Inference}
\label{sec:inference_not_use_constraint}
During inference, we use Eq.~(\ref{eq:select_index_with_argmin}) to quantize document embeddings instead of Eq.~(\ref{eq:ideal_posterior_distrib_comput}) or Eq.~(\ref{eq:approximate_solve_posterior_distrib_comput}). In this way, we can quantize each document embedding online efficiently. Otherwise, computing with Eq.~(\ref{eq:ideal_posterior_distrib_comput}) is expensive and Eq.~(\ref{eq:approximate_solve_posterior_distrib_comput}) introduces stochastic noise when batching documents.

\section{Experimental Setup}
Here we present our experimental settings, including datasets, baselines, and implementation details.

\subsection{Datasets and Metrics}

We conduct experiments on two large-scale ad-hoc retrieval benchmarks from the TREC 2019 Deep Learning Track~\cite{craswell2020overview, bajaj2016ms}, passage ranking and document ranking. They have been widely-adopted in previous works related to neural ranking. The passage ranking task has a corpus of $8.8$M passages, $0.5M$ training queries, $7k$ development queries~(henceforth, MARCO Passage), and $43$ test queries~(DL Passage). The document ranking task has a corpus of $3.2M$ documents, $0.4M$ training queries, $5k$ development queries~(MARCO Doc), and $43$ test queries~(DL Doc). For both tasks, we report the official metrics and R@100 based on the full-corpus retrieval results. 

\subsection{Baselines}
We exploit two types of baselines, vector compression methods and retrieval models.
 
\subsubsection{Vector Compression Baselines} \mbox{}

Unsupervised methods include PQ~\cite{jegou2010product}, ScaNN~\cite{guo2020accelerating}, ITQ+LSH~\cite{gong2012iterative}, OPQ~\cite{ge2013optimized}, and OPQ+ScaNN. We use Faiss library~\cite{johnson2019billion} to implement those baselines except for ScaNN~\cite{guo2020accelerating}, which is implemented based on its open-sourced code. 

Supervised methods include recently proposed DPQ~\cite{chen2020differentiable,zhang2021joint} and JPQ~\cite{zhan2021jointly}, both of which are also based on PQ. 
We re-implement DPQ since it is originally designed for word embedding compression~\cite{chen2020differentiable} and recommendation systems~\cite{zhang2021joint}. We use the same warmup process as RepCONC. JPQ is lately proposed for document ranking and shares the same warmup process.
Another compression method, BPR~\cite{yamada2021bpr} binarizes dense vectors and thus is limited to a fixed compression ratio~(32x). As RepCONC already achieves very small performance loss with a 64x compression ratio, we do not implement BPR for comparison. 


\begin{table*}
\centering
\robustify\bfseries
\caption{
Comparison with different compression methods on TREC 2019 Deep Learning Track. 
Compression ratio is set to 64x, i.e., 48 bytes per passage/document. 
*/** denotes that RepCONC performs significantly better than baselines at $p < 0.05/0.01$ level using the two-tailed pairwise t-test. `Unsup. Compr.' and `Sup. Compr.' denote unsupervised compression methods and supervised compression methods, respectively. 
Best compression method in each column is marked bold. 
}
\label{tab:results_compress}
\begin{threeparttable}	
\begin{tabular}{@{}ll
!{\color{lightgray}\vrule}
S[table-format=2, table-column-width=12mm]<{x}!{\color{lightgray}\vrule}
S[table-format=1.3,table-column-width=12mm]S[table-format=1.3,table-column-width=12mm]S[table-format=1.3,table-column-width=12mm]S[table-format=1.3,table-column-width=12mm]
!{\color{lightgray}\vrule}
S[table-format=1.3,table-column-width=12mm]S[table-format=1.3,table-column-width=12mm]S[table-format=1.3,table-column-width=12mm]S[table-format=1.3,table-column-width=12mm]@{}}
\toprule
\textbf{} & \multirow{2}{*}{\textbf{Model}} & \mc{\textbf{Compr.}}
& \multicolumn{2}{c}{\textbf{MARCO Passage}} & \multicolumn{2}{c!{\color{lightgray}\vrule}}{\textbf{DL Passage}} 
& \multicolumn{2}{c}{\textbf{MARCO Doc}} & \multicolumn{2}{c}{\textbf{DL Doc}} \\
 &  & \mc{Ratio}
 & {MRR@10} & {R@100} & {NDCG@10} & {R@100} 
 & {MRR@100} & {R@100} & {NDCG@10} & {R@100} \\ \midrule
 \multicolumn{2}{l}{\textbf{Uncompressed}} \\
 & ANCE~\cite{xiong2021approximate} & 1 & 0.338 & 0.862 & 0.654 & 0.445 & 0.377** & 0.894** & 0.610 & 0.273 \\ 
 & ADORE~\cite{zhan2021optimizing} & 1 & 0.347 & 0.876 & 0.683 & 0.473 & 0.405 & 0.919 & 0.628 		  & 0.317 \\ 
 \multicolumn{2}{l}{\textbf{Unsup. Compr. }} \\
 & PQ~\cite{jegou2010product} & 64 & 0.028** & 0.193** & 0.077** & 0.067** & 0.038** & 0.205** & 0.076** & 0.043**\\
 & ScaNN~\cite{guo2020accelerating} & 64 & 0.034 ** & 0.402** & 0.085** & 0.121** & 0.149** & 0.563** & 0.318** & 0.137**\\
 & ITQ+LSH~\cite{gong2012iterative} & 64 & 0.271 ** & 0.782** & 0.501** & 0.367** & 0.322** & 0.845** & 0.543** & 0.263**\\
 & OPQ~\cite{ge2013optimized} & 64 & 0.290** & 0.830** & 0.591** & 0.417** & 0.340** & 0.880** & 0.579 & 0.282**\\
 & OPQ+ScaNN & 64 & 0.310** & 0.837** & 0.586** & 0.429** & 0.357** & 0.893** & 0.564* & 0.286 \\
 \multicolumn{2}{l}{\textbf{Sup. Compr. }} \\
 & DPQ~\cite{chen2020differentiable,zhang2021joint} & 64 & 0.305** & 0.840** & 0.589** & 0.440 & 0.353** & 0.891** & 0.576 & 0.302\\
 & JPQ~\cite{zhan2021jointly} & 64 & 0.332** & 0.863 & 0.644 & 0.447 & 0.384** & 0.905* & \bfseries 0.608 & 0.302 \\ 
 & RepCONC~(Ours) & 64 & \bfseries 0.340 & \bfseries 0.864 & \bfseries 0.668 & \bfseries 0.492 & \bfseries 0.399 & \bfseries 0.911 & 0.600 & \bfseries 0.305 \\
\bottomrule
\end{tabular}
\end{threeparttable}
\end{table*}


\subsubsection{Retrieval Models} \mbox{}

First-stage retrieval models involve BoW models and DR models. 
BoW models include BM25~\cite{robertson1994some} and its variants, such as DeepCT~\cite{dai2019context}, HDCT~\cite{Dai2020ContextAwareDT}, doc2query~\cite{nogueira2019document}, and docT5query~\cite{nogueira2019doc2query}. 
DR models include RepBERT~\cite{zhan2020repbert}, ANCE~\cite{xiong2021approximate}, STAR~\cite{zhan2021optimizing}, and ADORE~\cite{zhan2021optimizing}. Their output embeddings are of dimension $768$ and are not compressed. All of them utilize negative sampling methods for training as RepCONC. In our experiments related to time efficiency, we use IVF~\cite{jegou2010product} with the same hyperparameters as RepCONC to accelerate ADORE~\cite{zhan2021optimizing}, the most competitive uncompressed DR baseline. 

Although several ranking models also conduct end-to-end retrieval, their latency is significantly higher than typical first-stage retrievers. Therefore, we classify them as complex end-to-end retrieval models. These models include ColBERT~\cite{Khattab2020ColBERTEA}, COIL~\cite{gao2021coil}, uniCOIL~\cite{lin2021few}, and DeepImpact~\cite{deepimpact}~\footnote{
Although uniCOIL~\cite{lin2021few} and DeepImpact~\cite{deepimpact} can leverage the inverted indexes like BM25~\cite{robertson1994some}, they are much slower possibly due to much smaller vocabulary size~(30k vs. 500k) and not removing stop words. 
}. 
Note, for COIL~\cite{gao2021coil}, the authors uploaded a new model trained with hard negatives in the github repository, which is not included in its paper. We denote it as COIL-Hard.

\subsection{Implementation Details}

Here are our model settings.
We build RepCONC based on huggingface transformers~\cite{wolf2019huggingface} and Faiss ANNS library~\cite{johnson2019billion}. The dual-encoders use RoBERTa-base~\cite{liu2019roberta} as the backbone, and the output embedding dimension is 768. Embedding similarity is computed with inner product. For PQ hyper-parameters, $K$ is set to 256, and $M$ is set to 4, 8, 12, 16, 24, 32, and 48 for different compression ratios. The compression ratio equals $4\times 768 / M$ since one vector is compressed to $M$ bytes. 

Training settings are as follows. 
Most training hyper-parameters are kept the same in both datasets except for batch size due to the limitation of GPU memory. 
Following ADORE~\cite{zhan2021optimizing}, training is in two stages. 
In the first stage where static hard negatives are used, the optimizer is AdamW~\cite{loshchilov2017decoupled}; learning rates are $5\times 10^{-6}$ and $2\times 10^{-4}$ separately for encoders and centroid embeddings; $\lambda$ in Eq.~(\ref{eq:weighted_sum_loss}) is set to 0.05 for $M=48/32/24$, 0.07 for $M=16$, 0.1 for $M=12$, 0.2 for $M=8$, and 0.3 for $M=4$; batch sizes are separately set to 1024 and 256 for passage and document ranking. 
In the second stage where dynamic hard negatives are used, the optimizer is AdamW~\cite{loshchilov2017decoupled}; learning rates are $2\times 10^{-6}$ and $2\times 10^{-5}$ separately for encoders and centroid embeddings; batch sizes are  set to 128. For $M=4$, we replace Eq.~(\ref{eq:ranking_oriented_loss}) with LambdaLoss~\cite{burges2010ranknet} for better ranking performance. Training time is about 4 hours for passage ranking and 2 hours for document ranking. 
 
Now we present our hardware settings and details about latency measurement. We use Xeon Gold 5218 CPUs and RTX 3090 GPUs. 
When training and measuring latency, we use one CPU thread and one GPU. Training time is about 9 hours for passage ranking and 2 days for document ranking on one RTX 3090 GPU. 
Additional notes about latency measurement are as follows. 
BoW search and vector search are both conducted on the CPU. For most neural retrieval models including RepCONC, query encoding is required and is performed on GPU. In our reranking experiments, the reranking models are also running on GPU.

\section{Experiments}
We empirically evaluate RepCONC to address the following three research questions:
\begin{itemize}
	\item \textbf{RQ1:} Can RepCONC substantially compress the index without significantly hurting retrieval effectiveness?
	\item \textbf{RQ2:} How does RepCONC perform compared with other retrieval models?
	\item \textbf{RQ3:} How does constrained clustering contribute to the effectiveness of RepCONC?
\end{itemize}

\subsection{Comparison with Compression Methods}

\begin{figure}
    \subfloat[MS MARCO Passage]{\label{}\includegraphics[width=0.85\linewidth]{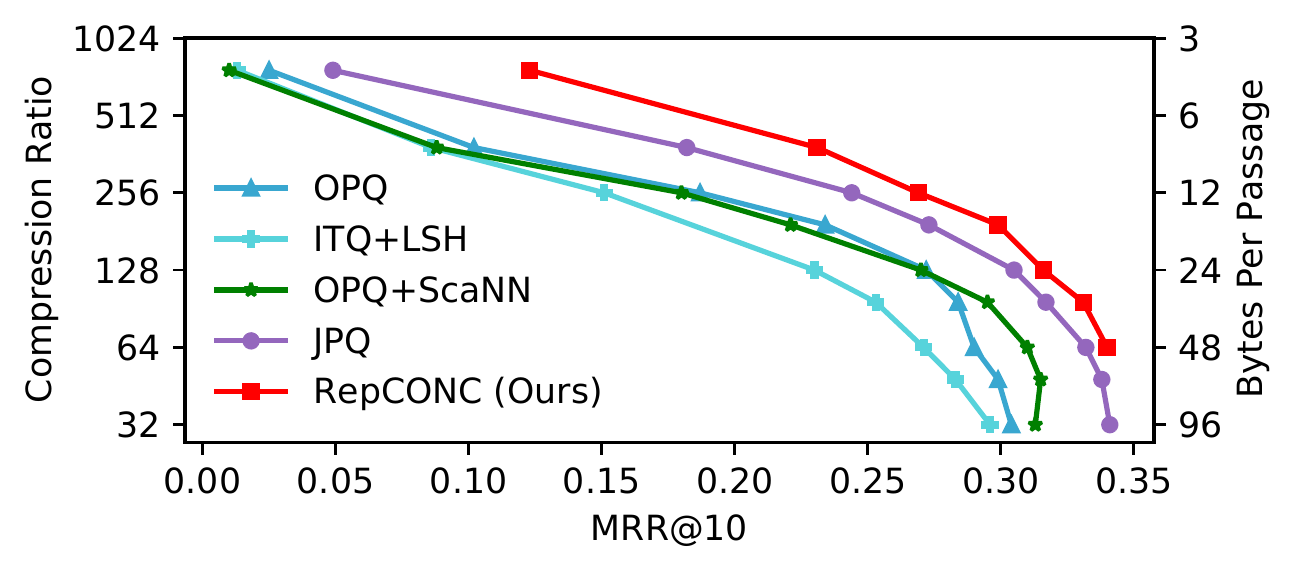}}  \\
    \subfloat[MS MARCO Document]{\label{}\includegraphics[width=0.85\linewidth]{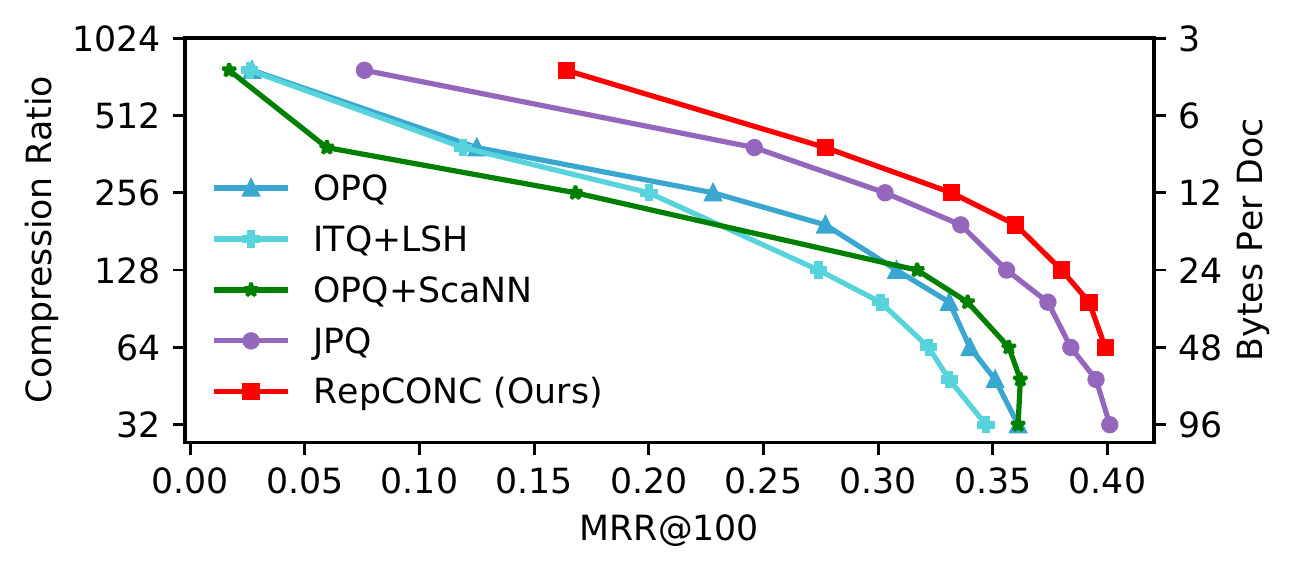}}
    \caption{Comparison with compression methods. Up and right is better.}
    \label{fig:compr_compress}
\end{figure}

This section compares RepCONC with vector compression baselines to answer \textbf{RQ1}. We compare it in two ways, a fixed 64x compression ratio and different compression ratios ranging from 64x to 784x. 

Ranking performances given a fixed 64x compression ratio are presented in Table~\ref{tab:results_compress}. Even if the index is compressed by 64 times, RepCONC outperforms ANCE~\cite{xiong2021approximate} and almost matches ADORE~\cite{zhan2021optimizing}, the state-of-the-art DR model trained by negative sampling. Compared with unsupervised compression methods, RepCONC exhibits significant performance gains and demonstrates the importance of joint learning. As for supervised compression baselines, RepCONC significantly outperforms DPQ~\cite{zhang2021joint, chen2020differentiable} and especially outperforms the recently proposed state-of-the-art JPQ model~\cite{zhan2021jointly} on most metrics. JPQ cannot train Index Assignments. It uses K-Means to generate them and fixes them during training. Our proposed RepCONC, on the contrary, is able to update Index Assignments during training and results show its effectiveness. 

Ranking performances in terms of different compression ratios are plotted in Figure~\ref{fig:compr_compress}. The advantage of RepCONC is more significant when larger compression ratios are used. 
For example, its MRR score is more than twice the JPQ's score when the compression ratio is 784x. 
We believe this is because RepCONC is able to generate high-quality Index Assignments specifically for ranking effectiveness, which becomes more important when fewer bytes are used. Instead, JPQ uses K-Means to produce task-blind Index Assignments and compromises ranking performance.

\subsection{Comparison with Retrieval Models}

This section compares RepCONC with various retrieval models to address \textbf{RQ2}. We firstly compare it with first-stage retrievers, including BoW models and DR models. Then we compare it with complex~(slow) end-to-end retrievers. 

\subsubsection{Comparison with First-Stage Retrievers} \mbox{}

\begin{figure}
    \subfloat[MS MARCO Passage]{\label{}\includegraphics[width=0.85\linewidth]{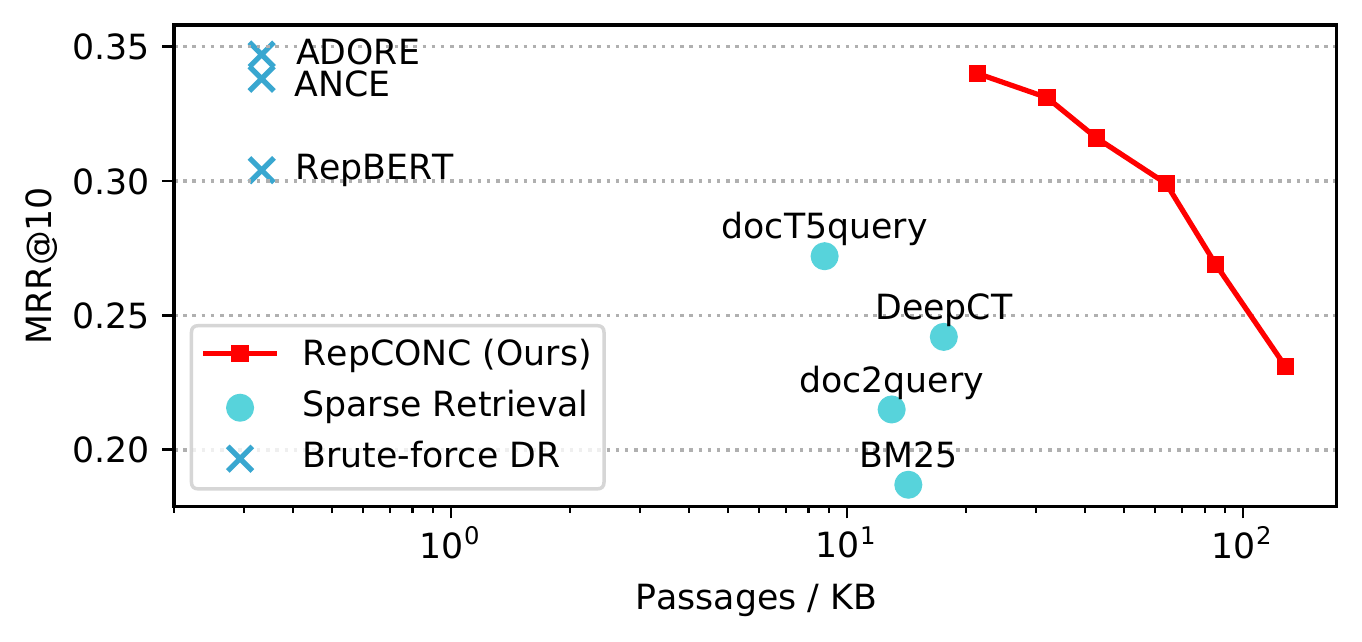}}  \\
    \subfloat[MS MARCO Document]{\label{}\includegraphics[width=0.85\linewidth]{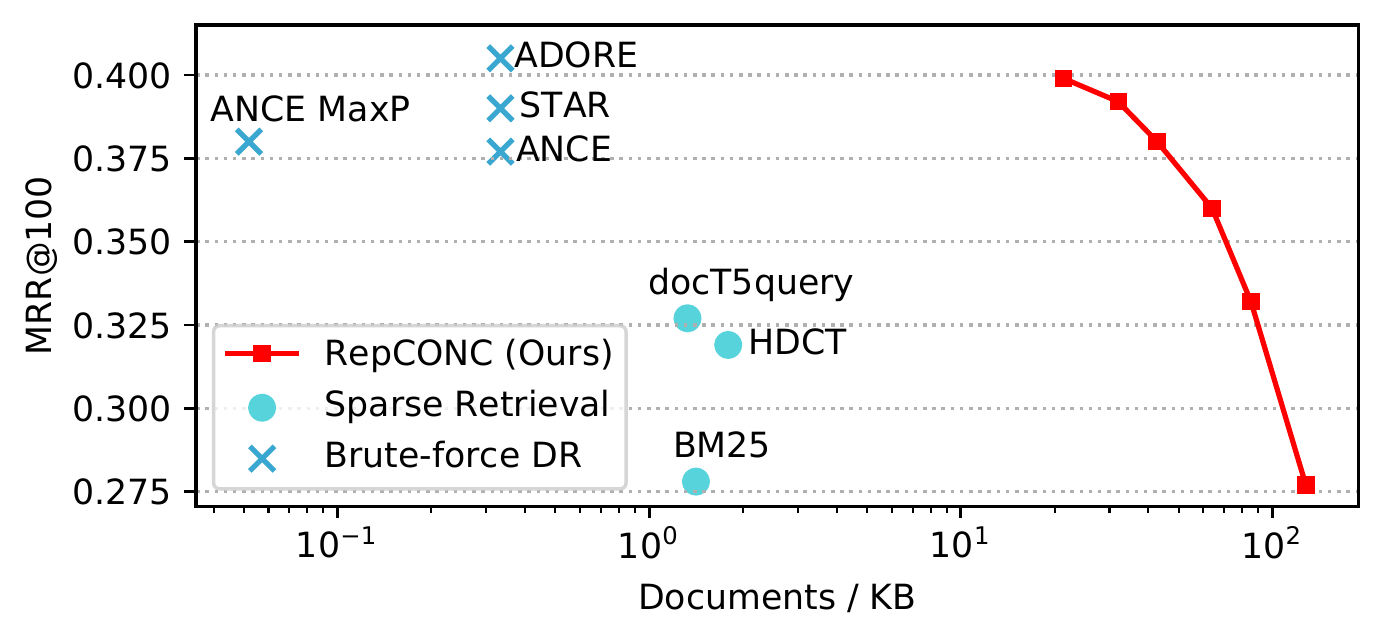}}
    \caption{Comparison with first-stage retrieval models in terms of effectiveness-memory trade-off. Up and right is better. The x-axis indicates the average number of passages/documents stored in 1 kilobyte. }
    \label{fig:compr_bowdr_mem}
\end{figure}

Figure~\ref{fig:compr_bowdr_mem} summarizes the effectiveness-memory tradeoff. As the figure shows, although DR models are much more effective than BoW models, they incur severe memory inefficiency. By jointly training the dual-encoders and quantization methods, RepCONC substantially improves memory efficiency of DR while still being very effective in ranking. It outperforms RepBERT~\cite{zhan2020repbert} and ANCE~\cite{xiong2021approximate} in effectiveness, and is almost as effective as ADORE~\cite{zhan2021optimizing}, the state-of-the-art DR model trained by negative sampling. Compared with BoW models, it can build a much smaller index, especially on document dataset where text is much longer than that on passage dataset. For example, on document ranking task, it can build a 100x smaller index than BM25 while still being equally effective.

\begin{figure}
    \subfloat[MS MARCO Passage]{\label{}\includegraphics[width=0.85\linewidth]{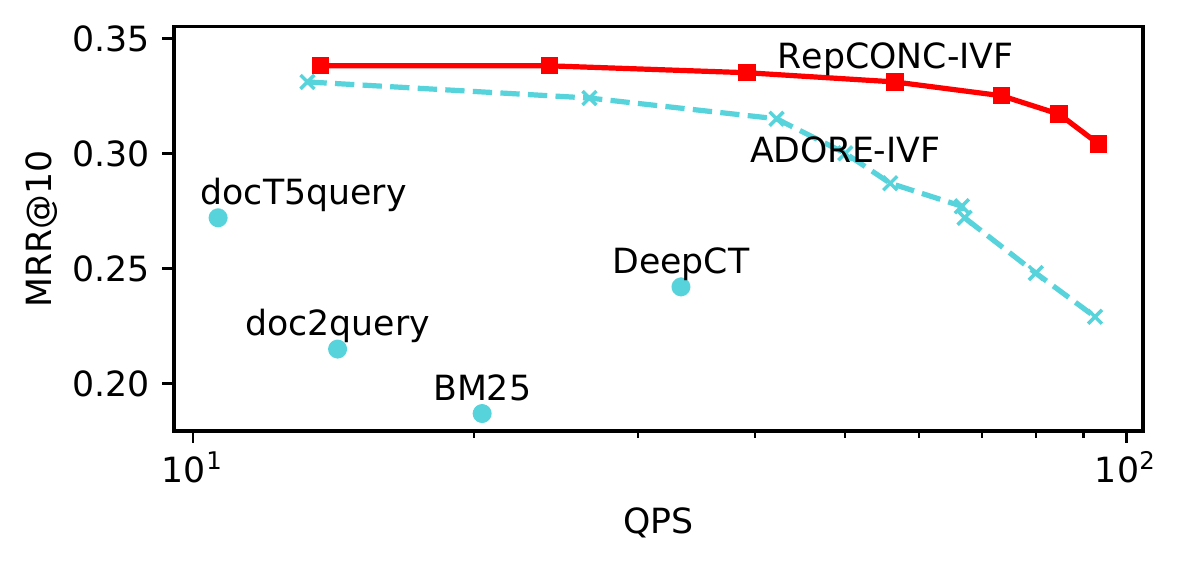}}  \\
    \subfloat[MS MARCO Document]{\label{}\includegraphics[width=0.85\linewidth]{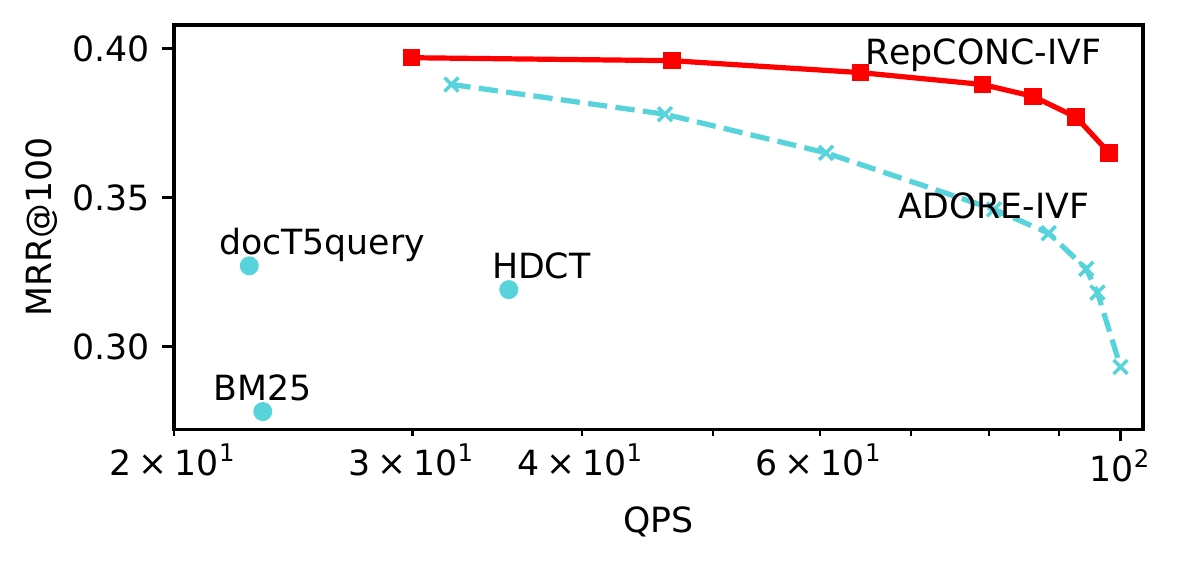}}
    \caption{Comparison with first-stage retrieval models in terms of effectiveness-latency trade-off. Up and right is better. The search is performed on CPU with one thread. QPS stands for `query per second'.}
    \label{fig:compr_bowdr_qps}
\end{figure}

Figure~\ref{fig:compr_bowdr_qps} summarizes the effectiveness-latency tradeoff. 
To verify that RepCONC is more time-efficient than existing uncompressed DR models, we arm the state-of-the-art uncompressed DR model, ADORE, with the same IVF method~\cite{jegou2010product} as RepCONC employs. 
As the figure shows, both RepCONC-IVF and ADORE-IVF substantially outperform BoW models with the help of IVF acceleration. Most importantly, RepCONC-IVF outperforms ADORE-IVF, especially at large QPS settings. This is because PQ already provides RepCONC with about 15x speedup compared with brute-force dense retrieval. Therefore, ADORE is more dependent on IVF than RepCONC and has to sacrifice more effectiveness for acceleration. The results demonstrate the time efficiency of RepCONC. 

\subsubsection{Comparison with Complex End-to-End Retrievers} \mbox{}

\begin{figure}
    \includegraphics[width=0.85\linewidth, keepaspectratio=True]{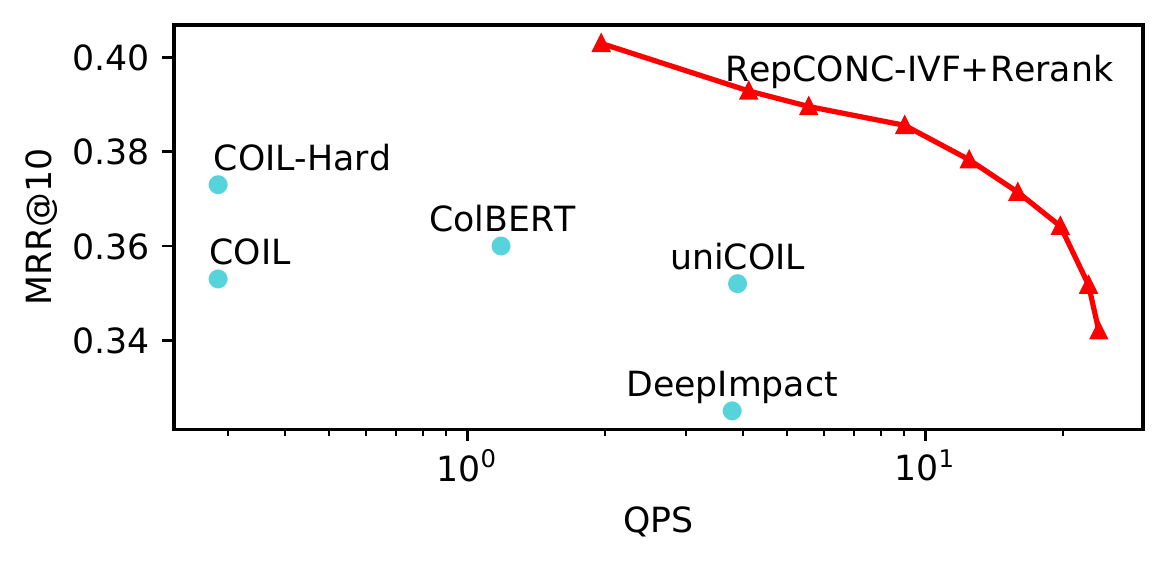}
    \caption{Comparison with complex~(slow) end-to-end retrieval models in terms of effectiveness-latency tradeoff on MS MARCO Passage Ranking. The search is performed on CPU with one thread. Up and right is better. QPS stands for `query per second'.
    } 
    \label{fig:compr_complex}
\end{figure}

This section compares RepCONC with some complex~(slow) end-to-end neural retrieval models. These models achieve better ranking performance with much higher query latency because of their complex model architecture. 
In consideration of fair comparison, we add a reranking stage to RepCONC and compare them in terms of effectiveness-latency tradeoff. The reranking models are MonoBERT and DuoT5 models open-sourced by the pygaggle library~\footnote{\url{https://github.com/castorini/pygaggle}}. MonoBERT firstly reranks top passages retrieved by RepCONC-IVF. Then DuoT5 further reranks the top passages output by MonoBERT. We tune the IVF speedup ratio, the MonoBERT reranking depth, and the DuoT5 reranking depth to evaluate ranking performance at different query latency. Note, query encoding and reranking are performed on GPU while the search is performed on CPU with one thread.

Ranking performances are summarized in Figure~\ref{fig:compr_complex}. We can see that RepCONC-IVF+Rerank substantially outperforms all baselines in terms of both effectiveness and time efficiency. In fact, RepCONC is also more memory-efficient than these baselines. The index size of RepCONC is less than 0.5GB, while COIL~\cite{gao2021coil}, ColBERT~\cite{Khattab2020ColBERTEA}, uniCOIL~\cite{lin2021few}, and DeepImpact~\cite{deepimpact} separately consumes 60GB, 162GB, 1.3GB, and 1.5GB for storing indexes.  
Therefore, RepCONC's efficient and effective retrieval is highly beneficial to second-stage reranking and helps the two-stage ranking achieve much better ranking performance than complex end-to-end retrieval models. 

\subsection{Ablation Study}

\begin{table}[t]
\caption{
Ablation study on MSMARCO Passage Ranking dataset. BPP stands for `bytes per passage'}
\label{tab:results_ablation}
\begin{tabular}{@{}ll
!{\color{lightgray}\vrule}
S[table-format=1.3,table-column-width=13mm]S[table-format=1.3,table-column-width=13mm]@{}}
\toprule
& \multirow{2}{*}{\textbf{Models}}  & \multicolumn{2}{c}{\textbf{MRR@10}}  \\ 
& & {\textbf{BPP:16}} & {\textbf{BPP:48}} \\ \midrule
\multicolumn{2}{l}{\textbf{Baselines}} \\
& DPQ~\cite{zhang2021joint, chen2020differentiable} & 0.244 & 0.305  \\
& JPQ~\cite{zhan2021jointly} & 0.273 & 0.332  \\
\multicolumn{2}{l}{\textbf{RepCONC}} \\
& OPQ~\cite{ge2013optimized} & 0.234 & 0.290  \\
& + Clustering & 0.275 & 0.332 \\
& + Constraint  & 0.284 & 0.337 \\
& + Dynamic Neg & 0.294 & 0.340 \\
\bottomrule
\end{tabular}
\end{table}

This section conducts an ablation study to answer \textbf{RQ3}. We summarize the results in Table~\ref{tab:results_ablation}. 

As the results show, clustering object helps RepCONC outperform DPQ~\cite{zhang2021joint, chen2020differentiable}. Although DPQ also utilizes a similar MSE loss, the gradients with respect to it only backpropagate to PQ centroids. Therefore, DPQ updates Index Assignments with a trick similar to Batch K-Means~\cite{bottou1995convergence} instead of clustering. However, Batch K-Means is shown to converge slowly~\cite{sculley2010web}. Besides, the target distribution of document embeddings is also changing during training, which makes convergence harder. 

With the help of the uniform clustering constraint, RepCONC outperforms the state-of-the-art JPQ method~\cite{zhan2021jointly}, which uses fixed Index Assignments generated by OPQ~\cite{ge2013optimized}. 
It demonstrates that simply adding a clustering loss is risky to retrieval effectiveness and that the constraint helps to tackle this problem by distinguishing the quantized vectors. The ranking performance is further improved by employing dynamic hard negatives~\cite{zhan2021optimizing}.

\begin{figure}
    \includegraphics[width=0.85\linewidth, keepaspectratio=True]{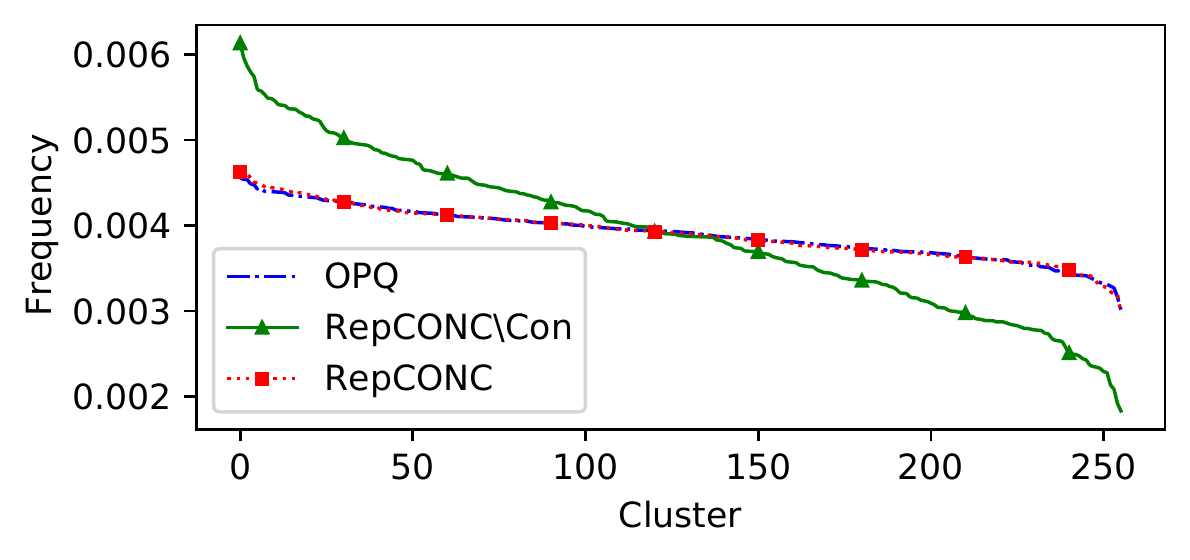}
    \caption{Cluster distribution on MS MARCO Passage Ranking. Clusters are sorted by the assigned frequency. Distributions across different sub-vector blocks are averaged. 
    RepCONC$\backslash$Con indicates RepCONC without constraint. 
    } 
    \label{fig:ablation_cluster_freq}
\end{figure}

To further verify that the constraint helps produce balanced clustering results, we plot the frequencies of clusters being assigned in Figure~\ref{fig:ablation_cluster_freq}. Without the constraint, RepCONC$\backslash$Con generates unbalanced clustering distribution. With the help of the constraint, the distribution is more balanced. It is similar to that of OPQ~\cite{ge2013optimized}, which uses K-Means for clustering. The distribution is not uniform because we do not use the constraint during inference as discussed in Section~\ref{sec:inference_not_use_constraint}.

\section{Conclusions}

To solve the efficiency issue existing in brute-force DR models, we present RepCONC, which learns discrete representations by modeling quantization as constrained clustering in the joint learning process. The clustering object requires the document embeddings to be clustered around the quantization centroids and facilitates joint optimization of PQ parameters and dual-encoders. To tackle the risk that clustering assigns vectors to only a few major centroids and results in indistinguishable quantized vectors, we introduce a uniform clustering constraint that enforces the vectors to be equally quantized to all possible centroids during training. The constraint is approximately solved as an instance of the optimal transport problem. In addition to constrained clustering, RepCONC employs the inverted file system~(IVF) to enable efficient vector search on CPUs. We conduct experiments on two widely-adopted ad-hoc retrieval benchmarks. Experimental results show that RepCONC significantly outperforms competitive quantization baselines and substantially improves the memory efficiency and time efficiency of DR. It substantially outperforms various retrieval models in terms of retrieval effectiveness, memory efficiency, and time efficiency. The ablation study demonstrates that constrained clustering is the key to the effectiveness of RepCONC.



\bibliographystyle{ACM-Reference-Format}
\balance
\bibliography{references}

\end{document}